# Time-Reversal Symmetry-Protected Coherent Control of Ultracold Molecular Collisions

Adrien Devolder[1], Timur V. Tscherbul[2], and Paul Brumer[1]*

[1]*Chemical Physics Theory Group, Department of Chemistry, and Center for Quantum Information and Quantum Control, University of Toronto, Toronto, Ontario, M5S 3H6, Canada*

[2]*Department of Physics, University of Nevada, Reno, NV, 89557, USA*

E-mail: adrien.devolder@utoronto.ca;ttscherbul@unr.edu,paul.brumer@utoronto.ca

**Abstract**

Coherent control of atomic and molecular scattering relies on the preparation of colliding particles in superpositions of internal states, establishing interfering pathways that can be used to tune the outcome of a scattering process. However, incoherent addition of different partial wave contributions to the integral cross sections (partial wave scrambling), commonly encountered in systems with complex collisional dynamics, poses a significant challenge, often limiting control. This work demonstrates that time-reversal symmetry can overcome these limitations by constraining the relative phases of S-matrix elements, thereby protecting coherent control against partial wave scrambling, even for collisions mediated by highly anisotropic interactions. Using the example of ultracold $O_2$-$O_2$ scattering, we show that coherent control is robust against short-range dynamical complexity. Furthermore, the time-reversal symmetry also protects the control against a distribution of collisional energy. These findings show that ultracold scattering into the final states that are time-reversal-invariant, such as the $J = 0$, $M = 0$ rotational state, can always be optimally controlled by using time-reversal-invariant initial superpositions. Beyond the ultracold regime, we observe significant differences in the controllability of crossed-molecular beam vs. trap experiments with the former being easier to control, emphasizing the cooperative role of time-reversal and permutation symmetries in maintaining control at any temperature. These results open new avenues for the coherent control of complex inelastic collisions and chemical reactions both in and outside of the ultracold regime.

# TOC Graphic

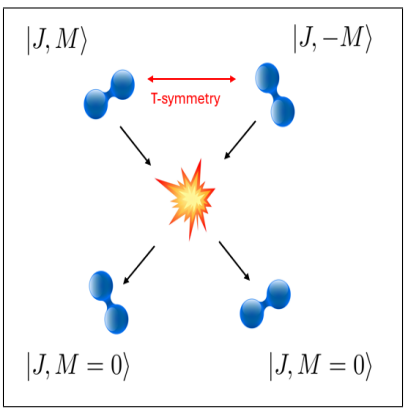

Coherent control of atomic and molecular scattering[1,2] is based on the initial preparation of colliding molecules in a superposition of internal states. This initial preparation creates different pathways that interfere with one another, altering scattering properties such as the cross section or the collisional rate. By varying the relative phase(s) in the initial superposition, these properties can be controlled. Nevertheless, coherent control of bimolecular collisions faces a major challenge due to the partial wave scrambling, which occurs because contributions from different initial and final partial waves add incoherently to the integral cross section.[3] Since the phases of the underlying scattering matrix elements are generally randomly distributed, the control landscape of each partial wave contribution varies, limiting the overall control of their sum, the experimentally observable integral cross section.[3]

Despite these challenges, coherent control of collisions has been shown to be effective in cold and ultracold regimes.[4,5] Complete control can be achieved for collisions occuring in the double s-wave regime (s-wave before and after the collision), such as in ultracold spin, charge or excitation exchange.[4] In this regime, partial wave scrambling does not occur. Additionally, complete control can be observed around an isolated resonance,[6] as typically only one partial wave contributes. However, control is lost when a large number of resonances overlap, as is the case in chaotic collisional dynamics.[7]

This complicated behavior is expected to be common in ultracold molecular collisions.[8–16] It is induced by a high density of rovibrational resonances at short-range, leading to the formation of a long-lived intermediate complex.[17] This complex can either collide with another molecule[18] or be excited by the trapping light,[19,20] resulting in losses from the trap and limiting the potential of ultracold molecular gases for quantum technologies. An effective solution is to prevent the molecules from approaching each other by using microwave[21–27] or electric shielding.[28] However, the short-range dynamics and their control can be of interest in many situations, such as in the study and control of ultracold chemical reactions. In such cases, alternative methods of control using either coherent superpositions[4] or external fields[29–31] must be considered. In this Letter, we demonstrate that time-reversal symmetry can protect control against partial wave scrambling, regardless of the complex collisional dynamics at short range mediated by highly anisotropic interactions.[32,33]

Since the beginning of research on the coherent control of scattering events, symmetries have played an important role. Time and space translation symmetries impose the superposition of states with the same energy and the same center-of-mass momentum, respectively.[1,2] Additionally, rotational symmetry requires the superposition of states with the same internal angular momentum projections for controlling the integral cross section[34] (but not the differential cross section[35]). Until now, symmetries have acted as constraints on the types of superpositions that can be used for coherent control. In this Letter, we demonstrate that symmetry can also aid in control by synchronizing the control of different partial waves and protecting it from anisotropic or chaotic collisional dynamics. As a result, the phase differences of the S-matrix elements for different partial waves are no longer randomly distributed but are constrained to fixed values by the symmetry.

*Coherent control of ultracold collisions with time-reversal superpositions* — Consider a collision between two molecules A and B, which initially are prepared in a coherent superposition of two internal angular momentum states $|j_A, m_A\rangle \otimes |j_B, m_B\rangle$:

$$|\Psi_{sup}\rangle = \cos\eta \, |j_A, m_{1A}\rangle \, |j_B, m_{2B}\rangle + \sin\eta \, e^{i\beta} \, |j_A, m_{2A}\rangle \, |j_B, m_{1B}\rangle, \quad (1)$$

where $m_{(1,2)A}$ and $m_{(1,2)B}$ are the projections of the internal angular momenta $j_A$ and $j_B$, respectively. $\eta \in [0, \pi/2]$ and $\beta \in [0, 2\pi]$ are the parameters that determine the relative population and phase of the superposition. For interference to occur, both states must have the same total internal projection, so $m_{1A} + m_{2B} = m_{2A} + m_{1B}$.[4,34] Note that the superposition (Eq. 1) is entangled. Although preparing entangled superpositions is more challenging experimentally, entangled pairs of CaF and KRb ultracold molecules have recently been created in the laboratory.[36–38] Non-entangled superpositions include two additional states, $|j_A, m_{1A}\rangle \, |j_B, m_{1B}\rangle$ and $|j_A, m_{2A}\rangle \, |j_B, m_{2B}\rangle$, which do not contribute to interference and are referred to as satellite terms.[1,5] The impact of satellite terms on the extent of control will be illustrated later for $O_2$-$O_2$ scattering. Since our goal is to demonstrate how time-reversal symmetry can enhance coherent control induced by quantum interference and protect control against partial wave scrambling, we will focus the following discussion on the entangled superposition (Eq. 1).

In ultracold collisions, only a single initial partial wave (the s-wave $\ell = 0$, $m_\ell = 0$) is in-volved. However, inelastic scattering and chemical reactions mediated by highly anisotropic interactions[29,32,33] can involve more than one final partial wave. Therefore, the state-to-state integral cross-section is given by:

$$\sigma_{sup \to f} = \frac{\pi}{k^2} \sum_{\ell', m'_\ell} \left| \cos(\eta) S_{m_{1A}, m_{2B}, 0, 0 \to f, \ell', m'_\ell} + \sin(\eta) e^{i\beta} S_{m_{2A}, m_{1A}, 0, 0 \to f, \ell', m'_\ell} \right|^2, \quad (2)$$

where $\ell'$ and $m'_l$ are the final partial wave and its projection on a quantization axis, and $S_{m_{1A}, m_{2B}, 0, 0 \to f, \ell', m'_\ell}$ is the S-matrix element between the initial state $|j_A, m_{1A}\rangle |j_B, m_{2B}\rangle$ in an s-wave ($\ell = 0$, $m_\ell = 0$) and the final state $|f\rangle$ in the final partial wave ($\ell', m'_l$). The incoherent addition of different partial wave contributions in Eq.(2) limits the control. For each partial wave, the optimized control parameters, $\eta$ and $\beta$, are different, which hinders the global optimization of $\sigma_{sup \to f}$.[3] As a consequence, the coherent nature of the control is quickly lost after the addition of even a few partial waves.[3]

Time-reversal symmetry can help to resolve this issue. If the angular momentum projections in Eq. (1) are chosen to have the opposite sign, $m_{1A} = -m_{2A} = m_{1B} = -m_{2B} = m$, the states $|j_A, m\rangle |j_B, -m\rangle$ and $|j_A, -m\rangle |j_B, m\rangle$ are related to each other by time-reversal symmetry. We define the time-reversal superpositions

$$|\Psi_{sup}\rangle = \cos \eta \, |j_A, m\rangle |j_B, -m\rangle + \sin \eta \, e^{i\beta} |j_A, -m\rangle |j_B, m\rangle. \quad (3)$$

These superposition have already been shown to enhance coherence times of trapped alkali-metal atoms[39,40] and quantum sensing with ultracold Dy atoms.[41] If the final state is invariant under the time-reversal symmetry (for example $|j_f = 0, m_f = 0\rangle$), $\hat{T}|f\rangle = |f\rangle$, the two S-matrix elements in Eq.(2) are constrained by the following relation:[42]

$$S_{m, -m, 0, 0 \to f, \ell', 0} = p S_{-m, m, 0, 0 \to f, \ell', 0}, \quad (4)$$

where $p = \pm 1$ is the parity of the initial state. Since parity is conserved during collisions in the absence of an external electric field, only even or odd values of $\ell'$ are possible. Therefore, the relationship between the S-matrix elements is always the same (they are either equal or of the opposite sign) for every final partial wave $\ell'$ involved in the collision. The control of different final partial wave contributions is thereby synchronised, resulting in complete control.

We illustrate this with the expression for the state-to-state integral cross section for the initial superposition (3) and a time-reversal invariant final state:

$$\sigma_{sup \to f} = \frac{\pi}{k^2} \sum_{\ell'} \left| \cos(\eta) S_{m,-m,0,0 \to f,\ell',0} + \sin(\eta) e^{i\beta} S_{-m,m,0,0 \to f,\ell',0} \right|^2, \qquad (5)$$

using eq.(4):

$$\sigma_{sup \to f} = \frac{\pi}{k^2} \left( \sum_{\ell'} |S_{i,0,0 \to f,\ell',0}|^2 \left| \cos(\eta) + p\sin(\eta) e^{i\beta} \right|^2 \right). \qquad (6)$$

Since the parity $p$ is determined by that of the initial internal state, the coherent control enabling term $\left| \cos(\eta) + p\sin(\eta) e^{i\beta} \right|^2$ is the same for every final partial wave and hence can be optimized with the same parameters of the initial coherent superposition (3):

- $\eta_{min} = \eta_{max} = \frac{\pi}{4}$

- $\beta_{min} = 0$ and $\beta_{max} = \pi$ if $p = -1$ or $\beta_{min} = \pi$ and $\beta_{max} = 0$ if $p = 1$.

Complete control of the collisional transition $sup \to f$ is achieved by varying the phase of the time-reversal superposition $\frac{1}{\sqrt{2}} \left( |j_A, m\rangle |j_B, -m\rangle + e^{i\beta} |j_A, -m\rangle |j_B, m\rangle \right)$. The maximum value is $2\frac{\pi}{k^2} \left( \sum_{\ell'} |S_{i,0,0 \to f,\ell',0}|^2 \right)$, while the minimum value of the cross-section is zero, i.e a consequence of complete destructive interference. Note that these optimal parameters remain the same for any collisional energy within the ultracold regime. As a result, the time-reversal symmetry also protects the control against a narrow distribution of collisional energy.

Remarkably, the relation (4) is a manifestation of time-reversal symmetric properties of the initial states and the final state, and does not depend on the details of collision dynamics. In other words, this relation must be fulfilled regardless of the form of the interparticle potential and the complexity of the dynamics at short range. This imples that the control is protected from partial wave scrambling by time-reversal symmetry even in the presence of highly anisotropic potentials or chaotic collisional dynamics at short-range which are common in ultracold chemistry.[29,32,33] We note that the partial wave phase-locking mechanism[43,44] has also been shown to synchronize the different partial wave contributions.[3] In that case, the synchronization arises from the characteristics of the interaction and the collision dynamics, rather than from the symmetries of the initial and final states.

It is important to note that the time-reversal protection is expected to be sensitive to magnetic fields, which are often used in ultracold experiments and can break time-reversal symmetry. Magnetic fields already pose a difficulty for coherent control by lifting the degeneracy between the $m$-states. The results of this article provide further arguments that coherent control experiments should be conducted at low magnetic fields.

*Protection against interaction potential anisotropy* — We illustrate this control with the scat-tering of two oxygen molecules $^{17}O_2$ in their vibronic ground state, characterized by a moderately anisotropic interparticle potential. The molecular states in the coupled angular momentum basis are written as $|N, S = 1, J, M_J\rangle$ (neglecting the hyperfine structure), where $J$ is the total internal angular momentum of the molecule. For $N = 0$, the $O_2$ molecules can occupy three different states, corresponding to different projections $M_J = -1, 0$, and 1. These three states are degenerate in the absence of magnetic fields. For $N = 2$, the states split in three groups with different energies depending on the value of $J$, $J = 1, 2$ and 3. Here, we consider the three rotationally excited states of the first group ($J = 1$): $|N = 2, S = 1, J = 1, M_J = -1\rangle$, $|2, 1, 1, 0\rangle$ and $|2, 1, 1, 1\rangle$. For the rest of the paper, we write $|g_{M_J}\rangle = |0, 1, 1, M_J\rangle$ and $|e_{M_J}\rangle = |2, 1, 1, M_J\rangle$. When considering the scattering of identical molecules, the states of the collision complex must satisfy permutation symmetry.[4,45] Applying this symmetrization, our $\ell = 0$ initial superposition take the form:

$$|a, b\rangle = \frac{1}{\sqrt{2(1 + \delta_{ab})}} \left[|a\rangle |b\rangle + |b\rangle |a\rangle\right]. \tag{7}$$

where $|a\rangle$ and $|b\rangle$ are internal molecular states, e.g $|g_{M_J}\rangle$ and $|e_{M_J}\rangle$. We calculated the S-matrix elements between these symmetrized two-molecules states using the rigorous coupled-channel methodology outlined in our previous work.[4,45]

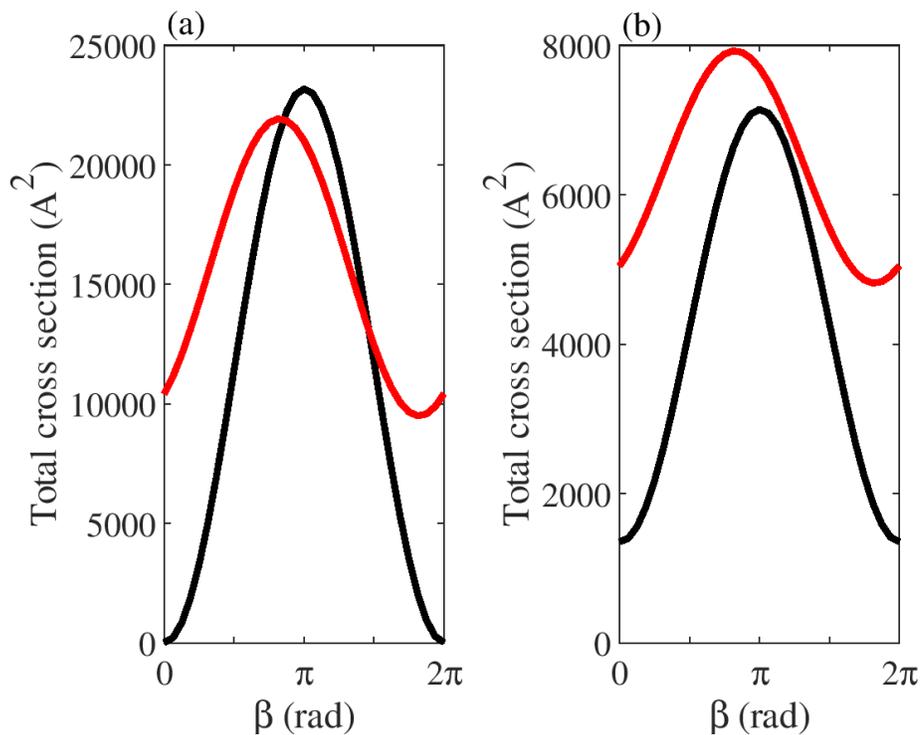

Figure 1: Coherent control of the integral cross section for ultracold $O_2$-$O_2$ collisions to the final states $|g_0, g_0\rangle$ (black) and $|g_0, g_{+1}\rangle$ (red).(a) The initial state is the entangled superposition (eq.8), while (b) the initial state is the non-entangled superposition (Eq.11)

The initial state is chosen to be the time-reversal superposition of the ground and rotationally excited states of $^{17}O_2$:

$$|\Psi_{sup}\rangle = \frac{1}{\sqrt{2}} \left(|g_{-1}, e_{+1}\rangle + e^{i\beta} |g_{+1}, e_{-1}\rangle\right). \tag{8}$$

We consider control of collisional de-excitation from this initial state to the final state $|g_0, g_0\rangle$, which transforms into itself under time-reversal symmetry. For non time-reversal invariant final states, other interesting features than the protection against partial wave scrambling are constrained by the time-reversal symmetry and are illustrated in Supplementary Information S1.

The control of the integral cross-section (ICS) from this superposition at 1 $\mu$K is shown in Figure 1 (a). We observe complete control over the ICS , with the minimum value of $1.9 \times 10^{-4}$ Å$^2$ and the maximum value of 23175 Å$^2$, a remarkable control range of over 9 orders of magnitude. The superposition $\frac{1}{\sqrt{2}} (|g_{-1}, e_{+1}\rangle + |g_{+1}, e_{-1}\rangle)$ minimizes the ICS while the superposition $\frac{1}{\sqrt{2}} (|g_{-1}, e_{+1}\rangle - |g_{+1}, e_{-1}\rangle)$ maximizes it, demonstrating that $p = -1$ in this case. For comparison, we also show the control for the final state $|g_0, g_{+1}\rangle$ in Figure 1 (a). This state is not invariant under time-reversal symmetry, and thus the control is not protected: the minimum value of the ICS is 9498 Å$^2$ while the maximum value is 21928 Å$^2$. Although the cross section can be varied by a factor of 2, this is significantly lower compared to the final state $|g_0, g_0\rangle$, which is time-reversal invariant. The achievement of complete control for the final state $|g_0, g_0\rangle$ is made possible by the synchronization of the control of all final partial waves, as illustrated in Figure 2(a), where the contributions of $\ell' = 0, 2$ and 4 exhibit the same control landscape. In constrast, for the final state $|g_0, g_{+1}\rangle$, the control profiles of the $\ell' = 2$ and 4 contributions are shifted relative to each other.

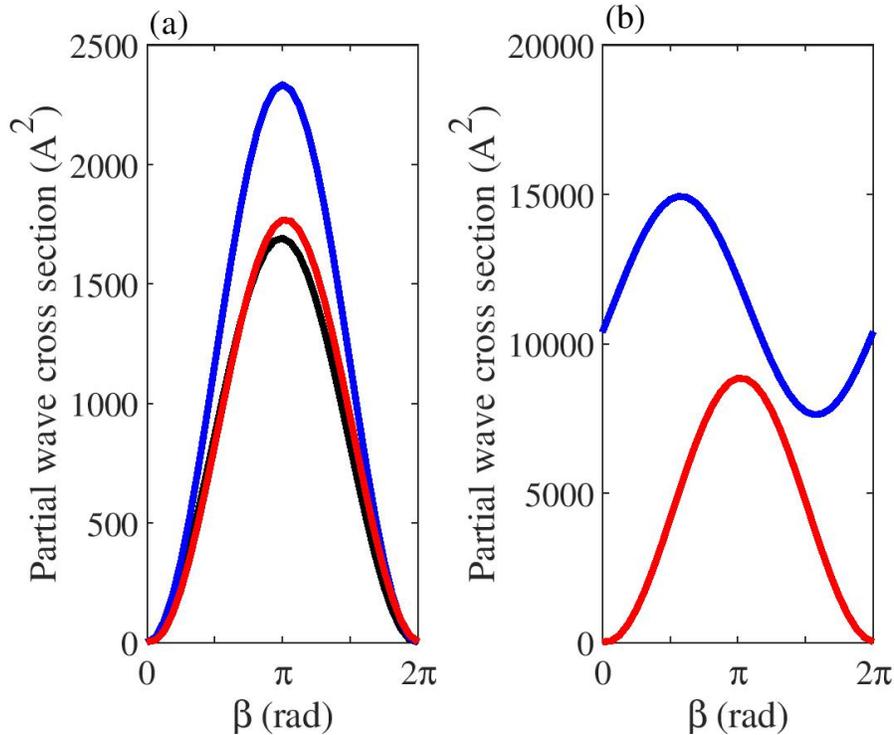

Figure 2: Coherent control of partial wave resolved cross sections to the final states (a) $|g_0, g_0\rangle$ and (b) $|g_0, g_{+1}\rangle$. The partial-wave resolved cross-section are plotted in the following way: s-wave (black), $d_0$-wave (red) and $g_0$-wave (blue)

The superposition $|\Psi_{sup}\rangle$ is entangled and can be experimentally challenging to create, even though recent experiments[36–38] showed that it is possible to do so with high fidelity. As an alternative, we consider the creation of time-reversal pair states of each of the molecules. More precisely, the first molecule is prepared in the superposition of $|g_{-1}\rangle$ and $|g_{+1}\rangle$ states, while the other molecule is prepared in the superposition of the $|e_{-1}\rangle$ and $|e_{+1}\rangle$ states:

$$|\psi_A\rangle = \frac{1}{\sqrt{2}} \left( |g_{-1}\rangle + e^{i\frac{\beta}{2}} |g_{+1}\rangle \right) \tag{9}$$

$$|\psi_B\rangle = \frac{1}{\sqrt{2}} \left( e^{i\frac{\beta}{2}} |e_{-1}\rangle + |e_{+1}\rangle \right), \tag{10}$$

The tensor product of these states must be symmetrized to account for the permutation symmetry of identical particles[45] giving

$$|\psi_2^S\rangle = \frac{1}{2}\left[|g_{-1}, e_{+1}\rangle + e^{i\beta}|g_{+1}, e_{-1}\rangle + e^{i\frac{\beta}{2}}(|g_{-1}, e_{-1}\rangle + |g_{+1}, e_{+1}\rangle)\right]. \tag{11}$$

This initial state can be created, for example, in merged beam experiments by separately preparing the molecules in coherent superpositions prior to collision.

In addition to the interfering states $|g_{-1}, e_{+1}\rangle$ and $|g_{+1}, e_{-1}\rangle$, there are then two other asymptotic two-molecule states $|g_{-1}, e_{-1}\rangle$ and $|g_{+1}, e_{+1}\rangle$ (satellite terms). The impact of these satellite terms on the control is shown in fig. 1 (b). The minimum value is 1352 Å$^2$, while the maximum value is 7142 Å$^2$. While complete destructive interference is lost due to the satellite terms, a large extent of control is still preserved. A reduction of control extent is also observed for the final state $|g_0, g_{+1}\rangle$ where the minimum and maximum values of the cross-section are 4817 Å$^2$ and 7925 Å$^2$, respectively. Note that in the double s-wave regime, complete control is achieved with non-entangled superpositions, due to the suppression of the satellite terms by the Wigner threshold laws.[4,46] This is generally not the case for ultracold exothermic inelastic scattering.

*Beyond the ultracold regime* — We have demonstrated the protection of coherent control by time-reversal symmetry in the ultracold regime. But what occurs beyond this regime? As shown below, this depends on the experimental conditions, such as those in crossed-molecular beam experiments versus trap/gas experiments. The key distinction between these two setups lies in the knowledge of the relative orientation of the incoming collision flux. In the former, this orientation is precisely known, while in the latter, it is undetermined and must be averaged over.

First we analyze the control in a crossed molecular beam experiment. If the z-axis of the laboratory frame is fixed along the direction of the initial relative momentum, we have $m_\ell = 0$ for every initial partial wave. Furthermore, since the internal projections of the time-reversal superposition (3) and the time-reversal invariant final state are both zero, the projection of the final partial waves is also fixed to zero. The expression for the integral cross section for time-reversal invariant final states $|f\rangle$ becomes:

$$\sigma_{sup \to f} = \frac{\pi}{k^2} \sum_{\ell'} \left| \sum_{\ell} \sqrt{2\ell+1} i^\ell \left( \cos\eta S_{m,-m,\ell,0 \to f,\ell',0} + \sin\eta e^{i\beta} S_{-m,m,\ell,0 \to f,\ell',0} \right) \right|^2 \quad (12)$$

Notably, the sum over the initial partial waves is inside of the squared magnitude, illustrating the presence of interference between initial partial waves due to the fixed orientation of the incident collision flux with respect to a laboratory-fixed quantity.[47]

To protect control against partial wave scrambling,[3] the time-reversal operation should impose a fixed phase relation between the S-matrix elements $S_{m,-m,\ell,0 \to f,\ell',0}$ and $S_{-m,m,\ell,0 \to f,\ell',0}$. However their relation includes a prefactor related to the parity of the initial partial waves :

$$S_{m,-m,\ell,0 \to f,\ell',0} = (-1)^\ell p S_{-m,m,\ell,0 \to f,\ell',0}. \quad (13)$$

This prefactor induces a phase shift of $\pi$ between the even and odd partial waves. Although it is not possible to achieve complete control over all partial waves, time-reversal symmetry enables the synchronous control of all partial waves of the same parity. Consequently, it becomes possible to suppress the contribution of either all even partial waves or all odd partial waves by tuning the relative phase $\beta$. This is illustrated for the final state $|g_{-1}, g_{+1}\rangle$ in Fig. 3 (a) and (c). Note that the final state $|g_{-1}, g_{+1}\rangle$ remains invariant under time-reversal symmetry due to the identical nature of the particles and the associated symmetrization (eq. 7).

This permutation symmetry can aid in achieving complete control when the two molecules occupy the same state after collisions, such as in $|g_0, g_0\rangle$. In this case, since the oxygen molecules $^{17}O_2$ are bosons, only even partial waves are allowed and their control is synchronised due to the time-reversal symmetry, as illustrated on Fig. 3 (b) and (d). Consequently, complete control over the partial wave contribution is achievable at any collision energy and temperature. However, this requires the cooperative effects of both time-reversal and permutation symmetries.

For experiments in a trap/gas, the initial orientation of the molecular incoming flux is unde-

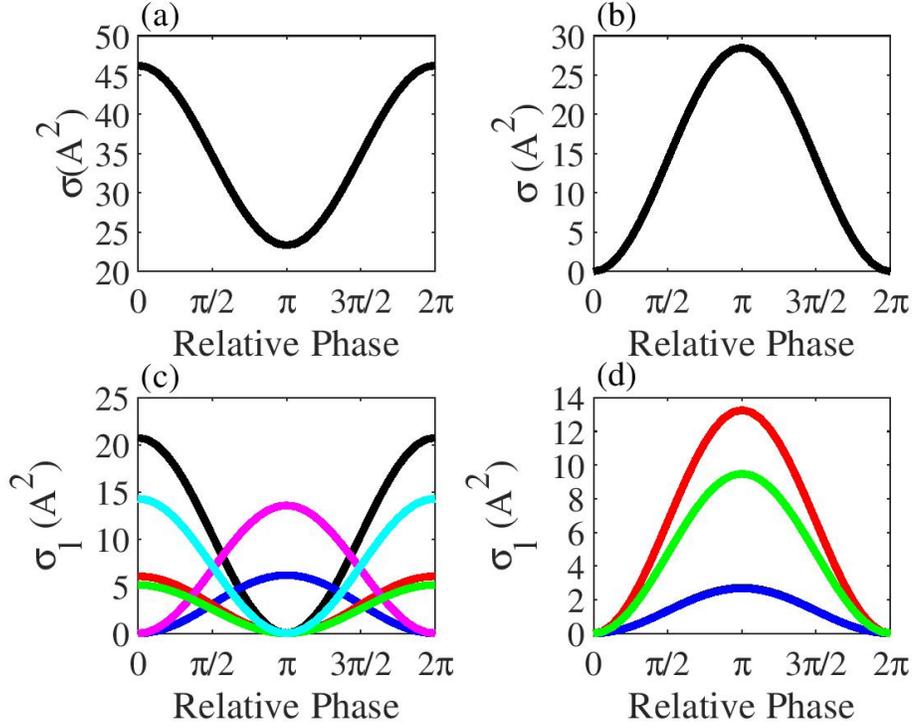

Figure 3: (a) and (c) Coherent control from the the time-reversal superposition (8) to the final states $|g_{-1}, g_{+1}\rangle$ for $O_2$-$O_2$ collisions in molecular-crossed beam experiment at 100 mK. Fig.(a) shows the total cross section while the final partial wave decomposition is shown on (c). The partial-wave resolved cross-section are plotted in the following way: $\ell' = 1$(black), $\ell' = 3$ (red), $\ell' = 5$ (green), $\ell' = 7$ (cyan), $\ell' = 2$(blue) and $\ell' = 4$(magenta).
(b) and (d) are the same for the final state $|g_0, g_0\rangle$. The partial-wave resolved cross-section are plotted in the following way: $\ell' = 2$(blue), $\ell' = 4$(red) and $\ell' = 6$(green).

termined and must be averaged over. The expression for the cross section becomes:

$$\sigma_{sup \to f} = \frac{\pi}{k^2} \sum_{\ell,m_\ell} \sum_{\ell'} \left| \cos(\eta) S_{m,-m,\ell,m_\ell \to f,\ell',m_\ell} + \sin(\eta) e^{i\beta} S_{-m,m,\ell,m_\ell \to f,\ell',m_\ell} \right|^2. \quad (14)$$

In this case, the average over the initial orientation requires that the sum over the initial partial waves is outside of the squared magnitude. As a consequence, there is no longer interference between different initial partial waves. In addition to the parity prefactor, time-reversal symmetry also reverses the sign of $m_\ell$, leading to the following relations between the S-matrix elements:

$$S_{m,-m,\ell,m_\ell \to f,\ell',m_\ell} = (-1)^\ell p S_{-m,m,\ell,-m_\ell \to f,\ell',-m_\ell}. \quad (15)$$

For these reasons, the different partial wave contributions are shifted relative to each other and the overall control of the cross section is no longer protected, as illustrated on the Fig. 4.

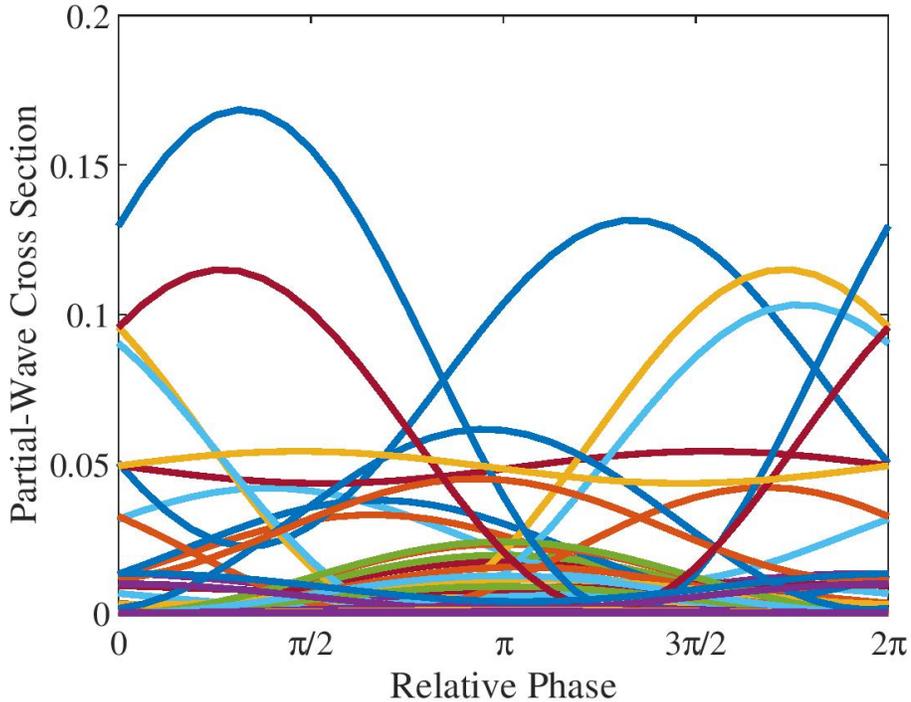

Figure 4: Coherent control of partial-wave resolved cross sections from the time-reversal superposition (8) to the final states $|g_0, g_0\rangle$ for $O_2$-$O_2$ collisions in trap/gas experiments at 100 mK

*Conclusion—* In the ultracold regime, time-reversal symmetry can protect coherent control against the partial wave scrambling induced by the complexity of the short-range collisional dy-namics, as demonstrated here for the moderately anisotropic $O_2$-$O_2$ scattering. Even though coupled-channel calculations can not yet be performed for chaotic collision dynamics between ultracold dialkali molecules,[16] the relation between the S-matrix elements imposed by the time-reversal symmetry [eq.(4)] should be valid. This suggests that using time-reversal superpositions, coherent control of ultracold molecular scattering into time reversal-invariant final states can be observed, even if the dynamics are chaotic and long-lived complexes are formed during the collisions. For example in the chemical reaction KRb+KRb $\rightarrow$ $K_2$+$Rb_2$, the preparation of KRb molecules in time-reversal initial superpositions can enable complete control over the population of the ground states of the products, $K_2$ ($J=0, M_J=0$) and $Rb_2$ ($J=0, M_J=0$).

Beyond the ultracold regime, a clear contrast emerges between crossed-molecular beam and trap experiments. In the latter, control is lost while in the former, time-reversal symmetry enforces synchronous control over final partial waves of the same parity. Moreover, we found that with identical particles, the permutation and time-reversal symmetry can cooperate to achieve synchronization of the partial wave contributions at any temperature.

**Acknowledgements:** This work was supported by the U.S. Office of Scientific Research (AFOSR) under contract number FA9550-22-1-0361. SciNet computational facilities are gratefully acknowledged.

# Time-Reversal Protected Coherent Control in Ultracold Molecular Collisions: Supplementary Information

Adrien Devolder[1], Timur V. Tscherbul[2], and Paul Brumer[1]*

[1]*Chemical Physics Theory Group, Department of Chemistry, and Center for Quantum Information and Quantum Control, University of Toronto, Toronto, Ontario, M5S 3H6, Canada*

[2]*Department of Physics, University of Nevada, Reno, NV, 89557, USA*

E-mail: adrien.devolder@utoronto.ca;ttscherbul@unr.edu,paul.brumer@utoronto.ca

# S1. Symmetric control landscape for non time-reversal invariant final states

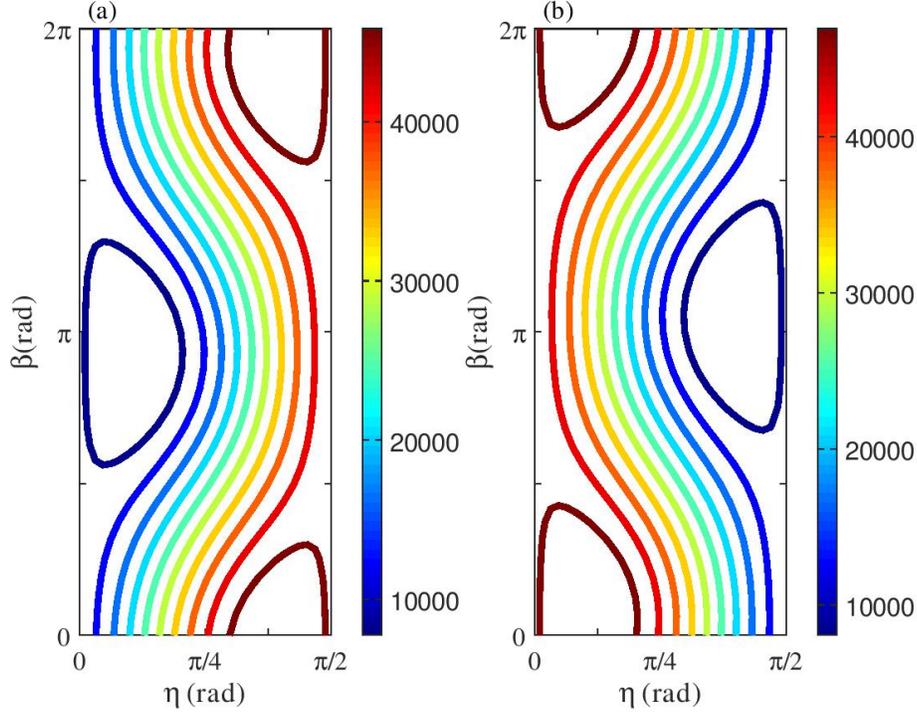

Figure 1: Coherent control of the integral cross-section for ultracold $O_2$-$O_2$ collisions from the entangled superposition $|\Psi_{ent}\rangle$ to (a) $|g_{-1}, g_{-1}\rangle$ or (b) $|g_{+1}, g_{+1}\rangle$

As shown in the main texts, time-reversal symmetry can protect control for final states that are invariant under it. However, even for other final states, using time-reversal superpositions can be advantageous. The first interesting property is the relation between control parameters for the final states $|f\rangle$ and $\hat{T}|f\rangle$ when they are different($|f\rangle \neq \hat{T}|f\rangle$). The cross-sections for the final states (that are time-reversal partners) $\sigma_{sup \to f}$ and $\sigma_{sup \to \hat{T}f}$ have the following forms:

$$\sigma_{sup \to f} = \frac{\pi}{k^2} \sum_{\ell'} \left| \cos(\eta) S_{m,-m,0,0 \to f,\ell',m'_\ell} + \sin(\eta) e^{i\beta} S_{-m,m,0,0 \to f,\ell',m'_\ell} \right|^2, \tag{1}$$

$$\sigma_{sup \to \hat{T}f} = \frac{\pi}{k^2} \sum_{\ell'} \left| \cos(\eta) S_{m,-m,0,0 \to \hat{T}f,\ell',-m'_\ell} + \sin(\eta) e^{i\beta} S_{-m,m,0,0 \to \hat{T}f,\ell',-m'_\ell} \right|^2, \tag{2}$$

where $m'_\ell = -m^f_{int} = m^{\hat{T}f}_{int}$, and $m^f_{int}$ and $m^{\hat{T}f}_{int}$ are the projection of the total internal angular momentum of the final state $|f\rangle$ and $\hat{T}|f\rangle$ respectively.

Due to the time-reversal symmetry, the following relations hold for the S-matrix elements: $S_{m,-m,0,0\to \hat{T}f,\ell',-m'_\ell} = pS_{-m,m,0,0\to f,\ell',m'_\ell}$ and $S_{-m,m,0,0\to \hat{T}f,\ell',-m'_\ell} = pS_{m,-m,0,0\to f,\ell',m'_\ell}$. Therefore, Eq.(2) becomes:

$$\sigma_{sup\to \hat{T}f} = \frac{\pi}{k^2} \sum_{\ell'} \left| \cos(\eta) S_{-m,m,0,0\to f,\ell',m'_\ell} + \sin(\eta)e^{i\beta} S_{m,-m,0,0\to f,\ell',m'_\ell} \right|^2. \tag{3}$$

Comparing Eqs. (1) and (3), we observe that they transform into each other under the transformation $\eta \to \pi/2 - \eta$ and $\beta \to 2\pi - \beta$. Therefore, the control landscapes for the $|f\rangle$ and $\hat{T}|f\rangle$ final states are related by a reflection with respect to $\eta = \pi/4$ and by a reflection with respect to $\beta = \pi$, i.e a point inversion with respect to $(\eta = \pi/4, \beta = \pi)$. We illustrate this with $|g_{-1}, g_{-1}\rangle$ as $|f\rangle$ and $|g_{+1}, g_{+1}\rangle$ as $|\hat{T}f\rangle$ in Figure 1. Furthermore, the following relationships hold between the minimum and maximum values of the control parameters:

$$\eta^{f}_{min} = \eta^{\hat{T}f}_{max}, \tag{4}$$

$$\beta^{f}_{min} = \pi - \beta^{\hat{T}f}_{max} = 2\pi - \beta^{\hat{T}f}_{min}, \tag{5}$$

where we have used the general relationship between the minimum and maximum values: $\eta^{f}_{min} + \eta^{f}_{max} = \pi/2$ and $\beta^{f}_{min} - \beta^{f}_{max} = \pi$ (and the same for $\hat{T}f$).

Another consequence of the time-reversal symmetry is the symmetric control landscape for the sum $\sigma_{sup\to f} + \sigma_{sup\to \hat{T}f}$ (see Fig. 2), since the time-reversal symmetry imposes that $\sigma_{m,-m\to f} + \sigma_{m,-m\to \hat{T}f} = \sigma_{-m,m\to \hat{T}f} + \sigma_{-m,m\to \hat{T}f}$. To demonstrate this, we start with Eqs. (1) and (3):

$$\begin{aligned}\sigma_{sup\to f} + \sigma_{sup\to \hat{T}f} &= \frac{\pi}{k^2} \sum_{\ell'} \left| \cos(\eta) S_{i,0,0\to f,\ell',m'_\ell} + \sin(\eta)e^{i\beta} S_{\hat{T}i,0,0\to f,\ell',m'_\ell} \right|^2 \\ &+ \frac{\pi}{k^2} \sum_{\ell'} \left| \cos(\eta) S_{\hat{T}i,0,0\to f,\ell',m'_\ell} + \sin(\eta)e^{i\beta} S_{i,0,0\to f,\ell',m'_\ell} \right|^2,\end{aligned} \tag{6}$$

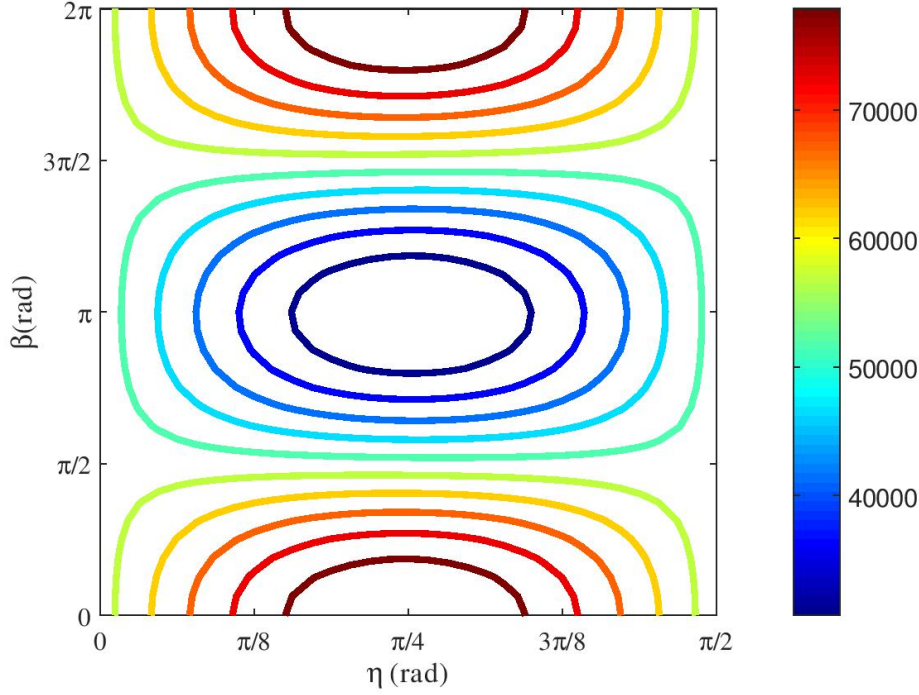

Figure 2: Coherent control of the ICS sum $\sigma_{sup\to f} + \sigma_{sup\to \hat{T}f}$ with $|f\rangle = |g_{-1}, g_{-1}\rangle$ and $|\hat{T}f\rangle = |g_{+1}, g_{+1}\rangle$ for ultracold $O_2$-$O_2$ collisions

Expanding the terms on the right-hand side, we find:

$$\sigma_{sup\to f} + \sigma_{sup\to \hat{T}f} = \frac{\pi}{k^2}\left[\sum_{\ell'}|S_{i,0,0\to f,\ell',m'_\ell}|^2 + |S_{\hat{T}i,0,0\to f,\ell',m'_\ell}|^2\right]$$
$$+ \frac{2\pi}{k^2}\cos(\eta)\sin(\eta)\left[\sum_{\ell'}|S_{i,0,0\to f,\ell',m'_\ell}||S_{\hat{T}i,0,0\to f,\ell',m'_\ell}|(\cos(\beta - \Delta_{\ell'}) + \cos(\beta + \Delta_{\ell'}))\right], \quad (7)$$

where $\Delta_{\ell'}$ is the difference of phases between the S-matrix elements $S_{i,0,0\to f,\ell',m'_\ell}$ and $S_{\hat{T}i,0,0\to f,\ell',m'_\ell}$. Using $\sigma_{i\to f} = \frac{\pi}{k^2}\sum_{\ell'}|S_{i,0,0\to f,\ell',m'_\ell}|^2$, $\sigma_{\hat{T}i\to f} = \frac{\pi}{k^2}\sum_{\ell'}|S_{\hat{T}i,0,0\to f,\ell',m'_\ell}|^2$, we finally obtain:

$$\sigma_{sup\to f} + \sigma_{sup\to \hat{T}f} = \sigma_{i\to f} + \sigma_{\hat{T}i\to f} + \frac{2\pi}{k^2}\cos(\eta)\sin(\eta)\left[\sum_{\ell'}|S_{i,0,0\to f,\ell',m'_\ell}||S_{\hat{T}i,0,0\to f,\ell',m'_\ell}|\cos(\Delta_{\ell'})\right]\cos(\beta). \quad (8)$$

Equation (8) is invariant under the transformation $\eta \to \pi/2 - \eta$ and $\beta \to 2\pi - \beta$. The control landscape is symmetric, as observed in Fig. 2. The minimum and maximum always occur at $\eta = \pi/4$. Their positions as a function of $\beta$ depend on the sign of the sum in the square bracket.

If the sum is positive, $\beta_{min} = \pi$ and $\beta_{max} = 0$. Otherwise, $\beta_{min} = 0$ and $\beta_{max} = \pi$.

More interestingly, this symmetric control landscape is also observed when summing over all final $m$-states. This occurs when the projection is not resolved during the measurement of the final states and therefore the measured cross section corresponds to a sum or an average over the projections. Time-reversal symmetry does not guarantee complete control for the $m$-summed cross sections but it does impose a symmetric control landscape. This symmetry of the control landscape demonstrates that the two interfering paths are indistinguishable, leading to maximal interference and control. Any change of basis would reduce the controllability as shown in,[1] so any other $m$-superposition would result in less effective control of the m-summed integral cross sections compared to the time-reversal superposition.